# Abnormal Surface Nonlinear Optical Responses in Topological Materials

Haowei Xu[1], Hua Wang[1], and Ju Li[1,2]

[1] Department of Nuclear Science and Engineering, Massachusetts Institute of Technology, Cambridge, Massachusetts 02139, USA

[2] Department of Materials Science and Engineering, Massachusetts Institute of Technology, Cambridge, Massachusetts 02139, USA

## Abstract

Nonlinear optical (NLO) responses of topological materials are under active research in recent years. Yet by far, most studies focused on the bulk properties, whereas the surface effects and the difference between surface and bulk responses have not been systematically studied. Here we develop a generic Green's function framework to investigate the surface NLO properties of topological materials. The Green's function framework can naturally incorporate many-body effects and can be easily extended to high-order NLO responses. Using $T_d$-$WTe_2$ as an example, we reveal that the surface can behave disparately from the bulk under light illumination. Remarkably, the shift and circular currents on the surface can flow in opposite directions to those in the bulk interior. Moreover, the light-induced spin current on the surface can be orders of magnitude stronger than its bulk counterpart. We also study the responses under inhomogeneous field and higher-order NLO effect, which are all distinct on the surface. These anomalous surface NLO responses suggest that light can be a valuable tool for probing the surface states of topological materials. On the other hand, the surface effects shall be prudently considered when investigating the optical properties of topological materials, especially if the material is of nanoscale and/or the light penetration depth is small.

# Introduction

In recent years, nonlinear optical (NLO) effects such as the bulk photovoltaic (BPV) effect have attracted substantial interest, owing to their potential applications in photodetection[1–4], energy harvesting[5–10], and material characterization[11–14]. The connection between topology and NLO properties is particularly intriguing. Certain NLO responses are closely related to the topological properties such as the Berry curvature and quantum metric tensors, thus the NLO effects can be utilized as a probe of these topological properties[11–13,15–17]. On the other hand, the topological nature can boost the NLO responses[2,18,19], thus the efficiency of applications such as photodetection could be enhanced by using topological materials.

Regarding NLO responses in topological materials, most works hitherto focus on the bulk responses[1–4,11–27], with only a few exceptions[28–32]. Moreover, there are very scarce works that study the surface and bulk responses together and investigate the difference between them. Indeed, even in normal materials, the surface and bulk responses can be substantially different. A typical example is that surface naturally breaks the inversion symmetry, hence even-order responses are always allowed on the surfaces, even if the bulk possesses inversion symmetry. In topological insulators, the bulk should be silent under light with frequencies below bulk bandgap, while the surface can be active due to the gapless topological surface states. These subtleties necessitate careful inspections of the surface effects. Typically, a slab model with finite thickness is used to study the surface effects. This approximate model omits a few essential interactions: the surface electrons should interact with all the bulk electrons, and the bulk should be infinite in depth. Besides these concerns pertinent to surface effects, previous studies on the NLO effects are mostly based on the independent particle approximation, whereas the many-body effects are ignored. Indeed, many-body effects such as electron-phonon coupling[33,34], excitonic effect[35,36] and strong correlations[37] may greatly influence the optical responses.

In this work, we develop a generic many-body framework for computing the NLO effects based on the Green's function formalism, which can naturally incorporate various many-body effects. Besides second-order response such as BPV, our Green's function formalism can be systematically and conveniently extended to handle higher-order NLO effects and inhomogeneous light fields. We apply our framework to the surface states of topological materials, whose Green's functions are obtained with the iterative Green's function (IGF) method[38,39]. This approach enables a rigorous treatment of the surface-bulk interactions. We take type-II Weyl semimetal $T_d$-$WTe_2$ [40,41] as an example. The bulk of $T_d$-$WTe_2$ is non-centrosymmetric and gapless, which is similar to the surface from a symmetry or bandgap point of view. However, the NLO responses on the surface are distinct from those in the bulk interior. Specifically, the BPV charge current on the surface and in the bulk can flow in opposite directions (Figure 1). This striking behavior demonstrates that surface effects in topological materials can be significant. Also, the surface bulk spin photovoltaic (BSPV)[42,43] conductivity is colossal and can be larger than its bulk counterpart by a factor of 10. Hence the surface of topological materials can be efficient platforms for spintronics applications. In addition, we show that the responses under inhomogeneous fields, and higher-order NLO effects are all distinct between the surface and the bulk. These anomalous responses on the surfaces indicate that the NLO effects can be utilized to not only probe, but also differentiate surface and bulk properties. On the other hand, the surface responses shall be prudently considered when investigating NLO properties of topological materials. This is particularly important when the penetration depth of the light is small, where the contribution from the surfaces can dominate the total responses, or if the material is thin, like less than $10^2$ monolayers (~50nm) thick. Given the semiconductor industry has generally moved to sub-10nm technology nodes, it is opportune time that such surface effects are addressed theoretically.

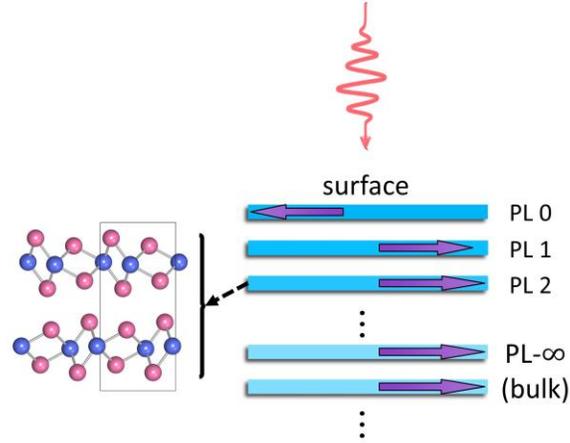

**Figure 1** A sketch of materials under light illumination. The surface and bulk principal layer (PL) corresponds to PL-0 and PL-∞, respectively. The optical responses of the surface and the bulk can be distinct (purple arrows). For $T_d$-$WTe_2$, each PL consists of two $WTe_2$ layer without inversion symmetry.

## Results

**General Theory.** To calculate the responses under light, one needs the thermodynamic and quantum average of the observable $\theta$, which can be formulated as[44]

$$\langle \theta \rangle = -i \int \frac{d^d\mathbf{k}}{(2\pi)^d} \frac{dE}{2\pi} \text{Tr}\{\theta G^<(\mathbf{k}, E)\} \tag{1}$$

Here $\theta$ can be a variety of observables, such as charge current ($\theta = -e\mathbf{v}$ with $\mathbf{v}$ as the velocity operator), spin current [$\theta = \frac{1}{2}(\mathbf{vs} + \mathbf{sv})$ with $\mathbf{s}$ as the spin operator], etc. $\int \frac{d^d\mathbf{k}}{(2\pi)^d}$ indicates the integration over the Brillouin zone in $d$-dimension, Tr is the trace operation. $G^<(\mathbf{k}, E)$ is the lesser Green's function and plays a role of "energy-resolved distribution function". In non-interacting systems, one has $[G_0^<(\mathbf{k}, E)]_{mn} = 2\pi i \delta_{mn} f_m \delta(E - E_m)$, where $m$ and $n$ are band indices, and $f_m$ and $E_m$ are the occupation number and energy of band $m$ at wavevector $\mathbf{k}$, respectively. In this case, one has $-i\int \frac{dE}{2\pi}[G_0^<(\mathbf{k}, E)]_{mn} = \delta_{mn} f_m$ and $\langle \theta \rangle = \int \frac{d^d\mathbf{k}}{(2\pi)^d} \sum_m \theta_{mm} f_m$, which is the usual thermal average of $\langle \theta \rangle$. When the electrons have interactions with phonon, defects, other electrons, etc., $G^<(\mathbf{k}, E)$ usually does not have a simple expression as in the non-interacting case, but it can be obtained using, e.g., Feynman diagrams[44]. In equilibrium, the expectation of certain observables, such as the charge current, should be zero. However, light can drive the system out of equilibrium, resulting in nonzero $\langle \theta \rangle$. Specifically, the interaction with light leads to a change in the less Green's function $\delta G^<$, which is dependent on the electric field $\mathcal{E}$. Then perturbatively one has $\langle \theta \rangle = A\mathcal{E} + B\mathcal{E}^2 + C\mathcal{E}^3 \cdots$, where $A$, $B$ and $C$ correspond to the first, second and third order optical responses and can be obtained from ab initio calculations (Methods).

In the following, we use WTe$_2$ in its T$_d$ phase as an example to study the optical response on the surface and in the bulk of topological materials. The unit cell of T$_d$-WTe$_2$ consists of two WTe$_2$ layers (Figure 1), which is used as a principal layer (PL). Each PL has interactions with other PLs, and these interactions can be included in the equilibrium Green's function $G_{PL}^0$, which is obtained using the IGF method[38,39]. Electrons on each PL also have interactions with phonons, defects, etc. These interactions are implicitly represented by a phenomenological electron lifetime $\tau$, which is taken to be a uniform value of 0.2 ps throughout this work. The Green's function of each PL from the surface into the bulk can be obtained with the IGF method, and the PL-resolved responses, defined as the responses localized on each PL, can be obtained by putting $G_{PL}^0$ in the

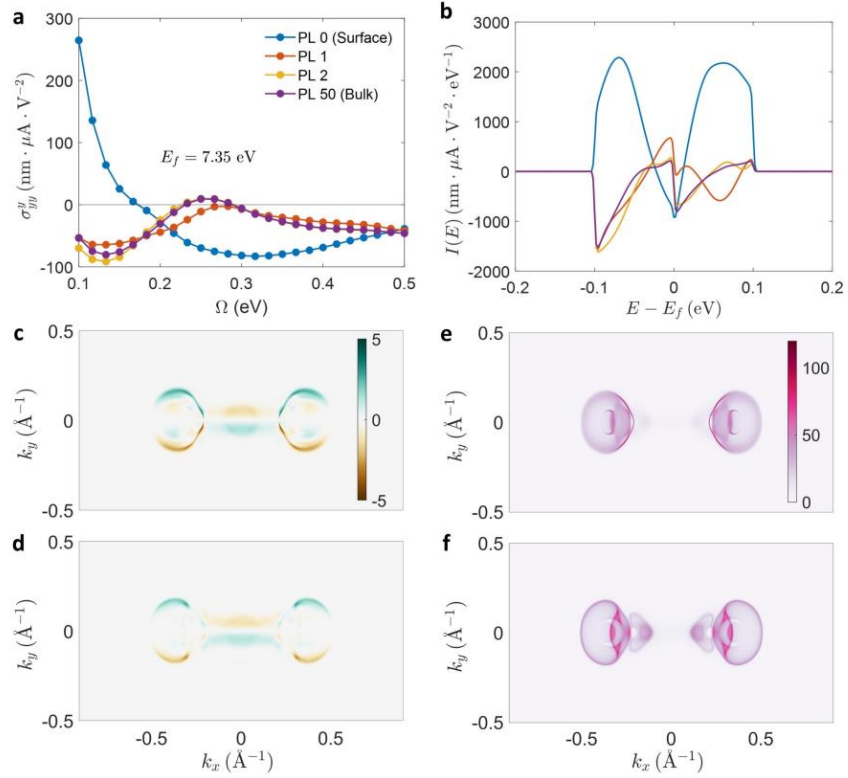

**Figure 2** BPV conductivity of $T_d$-$WTe_2$. **(a)** BPV conductivity localized on each PL from the surface to the bulk. **(b)** energy-resolve contribution to the BPV conductivity, defined as $I(E) \equiv -\frac{ie^3}{\omega^2 S}\sum_k Tr\{v^a G^<(E)\}$ for light frequency $\omega = 0.1\ eV$. **(c, d)** $k$-resolved contribution, defined as $I(k, E) \equiv -\frac{ie^3}{\omega^2 S} Tr\{v^a G^<(k, E)\}$ at $\omega = 0.1\ eV$ and $E = E_F$ for surface (c) PL and (d) bulk PL. **(e, f)** Spectrum function $A(k, E)$ at $E = E_F$ for surface (e) PL and (f) bulk PL.

Green's function formalism (Methods). Specifically, the surface and the bulk correspond to PL-0 and PL-∞, whose Green's functions are denoted as $G^0_{PL\text{-}0}$ and $G^0_{PL\text{-}\infty}$, respectively (Figure 1). In practice, $G^0_{PL}$ converges with PL ≳ 20, and we use $G^0_{PL\text{-}50}$ as the bulk Green's function ($G^0_{PL\text{-}\infty} \simeq G^0_{PL\text{-}50}$). The fermi level of $T_d$-$WTe_2$ is fixed at $E_F = 7.35$ eV [Figure S3 in Supplementary Materials (SM)] throughout this paper, and we study the optical responses in the mid-infrared regime ($\omega = 0.1 \sim 0.5$ eV).

**Bulk photovoltaic effect.** BPV effect indicates that in non-centrosymmetric materials, a DC charge current can be generated under light illumination without any external bias voltage. The BPV effect can be expressed as

$$j^a = \sigma^a_{bc}(0;\omega,-\omega)\mathcal{E}^b(\omega)\mathcal{E}^c(-\omega) \qquad (2)$$

Where $j^a$ is the current, while $a$ and $b/c$ are the directions of the current and the electric field, respectively. Here $\sigma^a_{bc}(0;\omega,-\omega)$ is the BPV conductivity and can be expressed as (SM)

$$\sigma^a_{bc}(0;\omega,-\omega) = -\frac{ie^3}{\omega^2 S}\sum_k \int \frac{dE}{2\pi} \text{Tr}\{v^a G^<(E)\} \qquad (3)$$

$$G^< = G_0^r(E) v^b G_0^r(E+\omega) v^c G_0^<(E)$$
$$+ G_0^r(E) v^b G_0^<(E+\omega) v^c G_0^a(E)$$
$$+ G_0^<(E) v^b G_0^a(E+\omega) v^c G_0^a(E)$$
$$+ (b \leftrightarrow c, +\omega \leftrightarrow -\omega)$$

One can see that the response operator $\theta$ in Eq. (1) has been taken as $-e\boldsymbol{v}$, which is the current density operator. $G_0^r, G_0^a$ and $G_0^<$ are the retarded, advanced and lesser Green's function of the system without light illumination, which are calculated with IGF. Here we assume that the light field is uniform. The $\boldsymbol{k}$ arguments in Eq. (3) are omitted for simplicity. $S$ is the area of the unit cell. $(b \leftrightarrow c, +\omega \leftrightarrow -\omega)$ indicates the simultaneous exchange of $b, c$ and $+\omega, -\omega$, which symmetrizes $\mathcal{E}^b$ and $\mathcal{E}^c$.

The PL-resolved BPV conductivities of $T_d$-WTe$_2$ under linear polarized light (shift current) with $y$-polarization are shown in Figure 2a. Obviously, the surface layer (PL-0) behaves very distinctly from the other interior PLs, whereas the first interior PL (PL-1) behaves similarly to the bulk. This implies that the surface effects mainly influence the outmost PL, whose thickness is about 1.4 nm. Remarkably, in some frequency region $\sigma_{\text{PL-0}}$ has even the opposite sign to $\sigma_{\text{PL-1}}$ and $\sigma_{\text{PL-}\infty}$, indicating that the under light illumination, the local charge current would flow in opposite directions on the surface PL. This is counter-intuitive, as PL-0 is directly attached to PL-1. Moreover, in some frequency regions $\sigma_{\text{PL-}\infty}$ is close to zero, while $\sigma_{\text{PL-0}}$ has a finite value, thus the current would flow mostly on the surface layer. The BPV conductivities under circularly polarized light (circular current) are also distinct for the surface and the bulk (SM Figure S4). The distinct BPV responses reveal that geometric and topological properties, which are closely related with BPV[15–17], are very distinct on the surface than from the bulk. Practically, this suggests that the layer resolved NLO responses can be harnessed to probe the surface states of topological materials. Another interesting observation is that the counter-propagating currents (Figure 1) may lead to a magnetic field **B** between PL-0 and PL-1, which can be estimated using Ampère–Maxwell law as $B \approx \frac{1}{2}\mu_0 j = \frac{1}{2}\mu_0 \sigma \mathcal{E}^2 \sim 6.3 \times 10^{-4} \mathcal{E}^2$ [T], where $\mu_0$ is the vacuum permittivity, $\sigma$ is taken as 100 nm · μA/V², and $\mathcal{E}$ is in the unit of MV/cm. Note that this magnetic field **B** has a pure orbital magnetic origin, as no spin contribution was counted. Specifically, a mild electric field of 5 MV/cm can generate an interlayer magnetic field of 16 mT, which can be detectable.

To better understand the surface effect, we look at the $(\mathbf{k}, E)$-resolved contribution to $\sigma$ in Eq. (3). The energy spectrum $I(E) \equiv -\frac{ie^3}{\omega^2 S}\sum_\mathbf{k} \text{Tr}\{v^a G^<(E)\}$ for $\omega = 0.1$ eV is plotted in Figure 2b, where one can see that $I_{\text{PL-0}}(E)$ is generally different from $I_{\text{PL-}\infty}(E)$. Besides, although $G^<(E)$ should be nonzero for a wide range of $E$, $I(E)$ is nonzero only when $|E| \lesssim \omega$. This indicates that the shift current is essentially a resonant interband process: light with frequency $\omega$ can assist the electrons to transit from $(\mathbf{k}, E)$ to $(\mathbf{k}, E \pm \omega)$, but due to the Pauli exclusion principle, such a transition is allowed only when $f_{\text{FD}}(E)[1 - f_{\text{FD}}(E \pm \omega)] > 0$, leading to $|E| \lesssim \omega$. Here $f_{\text{FD}}$ is the Fermi-Dirac distribution. The $\mathbf{k}$-resolved contribution defined as $I(\mathbf{k}, E) \equiv -\frac{ie^3}{\omega^2 S}\text{Tr}\{v^a G^<(\mathbf{k}, E)\}$ reveals more detailed information on the surface effects. $I(\mathbf{k}, E)$ for $\omega = 0.1$ eV at the Fermi level $E = E_F$ is plotted in Figure 2c-d. The difference between $I_{\text{PL-0}}(\mathbf{k}, E_F)$ and $I_{\text{PL-}\infty}(\mathbf{k}, E_F)$ is also significant, and the difference can be better appreciated by comparing with the spectrum function, which is defined as $A(\mathbf{k}, E) \equiv i\text{Tr}\{G_0^r(\mathbf{k}, E) - G_0^a(\mathbf{k}, E)\}$ (Figure 2e-f). The Fermi arc is clearly seen in $A_{\text{PL-0}}(\mathbf{k}, E_F)$ and is absent in $A_{\text{PL-}\infty}(\mathbf{k}, E_F)$. One can see that the difference between $I_{\text{PL-0}}(\mathbf{k}, E_F)$ and $I_{\text{PL-}\infty}(\mathbf{k}, E_F)$ largely lies in the region where the surface Fermi arc is located. This indicates that the difference between $\sigma_{\text{PL-0}}$ and $\sigma_{\text{PL-}\infty}$ is mainly a topological effect. Indeed, for topologically trivial materials, $\sigma_{\text{PL-0}}$ and $\sigma_{\text{PL-}\infty}$ are generally close to each other (SM Figure S7).

**Bulk spin photovoltaic effect.** Electrons have both charge and spin degrees of freedom. When electrons move under light illumination, their charge degree of freedom leads to a charge current, which is the BPV effect discussed in the previous section. Concurrently, the spin degree of freedom leads to a spin current, which is called the bulk spin photovoltaic (BSPV) effect[42,43]. BPV and BSPV are cousin processes and have similar physical origins, but very different symmetry-selection rules as spin is a pseudo-vector. This can be harnessed for pure spin current generation – that is, the spin current is allowed by symmetry, while the charge current is forbidden[42]. As each electron carries a charge of $e$ and spin of $\frac{\hbar}{2}$, one may expect that in the sense of equivalating $\frac{\hbar}{2} = |e|$, BPV and BSPV should have similar magnitude. However, on the surface of topological materials such as $T_d$-WTe$_2$, BSPV can be stronger than BPV by a factor of 10, as we will show below. This makes the surfaces of topological materials ideal platforms for spintronics applications.

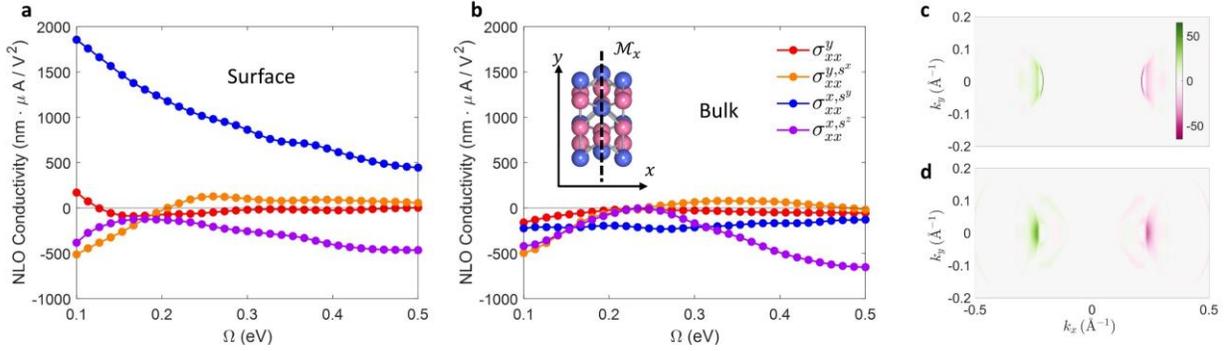

**Figure 3** BSPV conductivity for $T_d$-WTe$_2$. Different nonzero elements of the conductivity tensor are shown for the **(a)** surface PL and **(b)** bulk PL. **(c, d)** Spin density of states, defined as $D_S(\mathbf{k}, E) = iTr\{s[G_0^r(\mathbf{k}, E) - G_0^a(\mathbf{k}, E)]\}$ at $E = E_f$ for (c) surface PL and (d) bulk PL.

For spin current traveling in direction $a$ with spin polarization $i$, we set the response operator $\theta$ in Eq. (1) as $j^{a,s^i} = \frac{1}{2}(v^a s^i + s^i v^a)$, where $s^i$ is the spin operator. The BSPV conductivity $\sigma_{bc}^{a,s^i}(0; \omega, -\omega)$ has a similar expression to the BPV conductivity $\sigma_{bc}^a$ in Eq. (3) (SM Sec. 2.1). Note that we divide $\sigma_{bc}^{a,s^i}$ by $\frac{\hbar}{2e}$ so that it has the same unit as $\sigma_{bc}^a$. The mirror symmetry $\mathcal{M}_x$ of $T_d$-WTe$_2$ forbids some of the B(S)PV conductivity tensor, such as $\sigma_{xx}^x$ and $\sigma_{xx}^{x,s^x}$. This is because polar vectors such as $\mathcal{E}_x, v_x$ flip sign under $\mathcal{M}_x$, while axial vectors such as $s^x$ do not. Under linearly polarized light polarized in $x$-direction, the nonzero B(S)PV tensors are shown in Figure 3a-b (we do not consider the current along the out-of-plane $z$-direction). A prominent feature is that $\sigma_{xx}^{x,s^y}$ on the surface is almost ten times larger than its bulk counterpart and is also ten times larger than other components such as $\sigma_{xx}^y$. Indeed, hundreds of nm · μA/V$^2$ are typical values for B(S)PV conductivities in topological materials[27,42], while $\sigma_{xx}^{x,s^y}$ on PL-0 is as large as 2000 nm · μA/V$^2$ at $\omega = 0.1$ eV. This indicates that the spin current generation is exceptionally efficient on the surface of T$_d$-WTe$_2$, and the surface of topological materials can be ideal platforms for spintronics applications. Such a strong spin current comes partially from the Rashba spin-orbit coupling on the surfaces. In Figure 3c we plot the "spin density of states", defined as $D_S(k, E) \equiv i\mathrm{Tr}\{s[G_0^r(k,E) - G_0^a(k,E)]\}$ at $E = E_F$. One can see that spin-$y$ polarization is stronger around the surface Fermi arc. Actually, spin-$x$ also get amplified on the surfaces (SM Figure S5), but from our calculations $\sigma^{y,s^x}_{xx}$ has similar values for the surface and the bulk. This suggests that BSPV is not solely determined by the spin polarization, but requires synergy with other factors, such as band velocities.

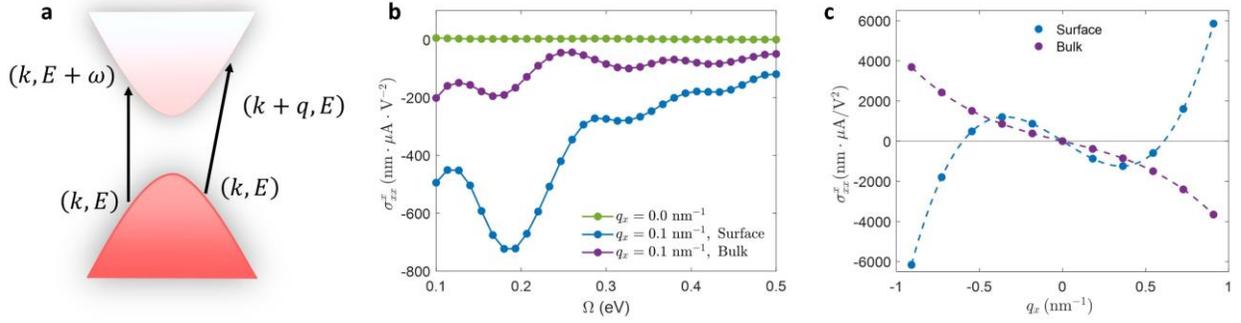

**Figure 4** BPV under inhomogeneous light field. **(a)** An illustration of the electron transitions under homogeneous (left arrow) and inhomogeneous (right arrow) field. **(b)** BPV conductivity under homogeneous ($q = 0$) and in inhomogeneous ($q_x \neq 0$) field. **(c)** The relationship between $\sigma^x_{xx}(\omega = 0.1\ eV; q)$ and $q_x$ for surface (blue curve) and bulk (purple curve) PL. The dashed curves are fittings of the solid dots with cubic functions.

**Inhomogeneous light field.** In the visible or infrared range, the photon wavevector $q$ is much smaller than the size of the Brillouin zone, thus one usually sets $q = 0$ when studying light-matter interactions. In other words, the light field is assumed to be homogeneous. However, the electromagnetic wave can be strongly inhomogeneous in the case of e.g., the Laguerre–Gaussian beam[45] with nonzero angular momentum. Subwavelength-scale variation also can be induced by plasmonic, polaritonic interactions[46,47]. In these situations, the spatial variation of the light field is strong and $q$ shall not be neglected[48]. In addition to these practical considerations, finite $q$ is also conceptually important, as it breaks certain spatial symmetries, and thus fundamentally alters the selection rules on optical process.

Our Green's function formalism provides a convenient way for incorporating the finite $q$ effect. Here we take the second-order BPV as an example. Under an inhomogeneous and oscillating electric field $\mathcal{E}^b(\omega, \boldsymbol{q})$ with frequency $\omega$ and wavevector $\boldsymbol{q}$, one still has a homogeneous and static current, which is

$$j^a = \sigma_{bc}^a(0; \omega, -\omega; \boldsymbol{q}, -\boldsymbol{q})\mathcal{E}^b(\omega, \boldsymbol{q})\mathcal{E}^c(-\omega, -\boldsymbol{q}) \tag{2}$$

where $\sigma_{bc}^a(0; \omega, -\omega; \boldsymbol{q}, -\boldsymbol{q})$ is the conductivity and can be calculated with the Green's function formalism (SM Sec. 2.2). We will use $\sigma_{bc}^a(\omega, \boldsymbol{q}) \equiv \sigma_{bc}^a(0; \omega, -\omega; \boldsymbol{q}, -\boldsymbol{q})$ as a shorthand. Intuitively, an electron in state $(\boldsymbol{k}, E)$ can (virtually) absorbs momentum $\boldsymbol{q}$ and energy $\omega$ from the inhomogeneous light field, jump to a different state $(\boldsymbol{k} + \boldsymbol{q}, E + \omega)$, then return $\boldsymbol{q}$ and $\omega$ to the light field and jump back to its original state (Figure 4a). During this process the electron may displace in real space, resulting in a current. Notably, $\boldsymbol{q}$ identifies a preferred (or unique) direction for such process, which could break certain spatial symmetries, including inversion, mirror, and rotation symmetry.

T$_d$-WTe$_2$ has mirror symmetry $\mathcal{M}_x$. As a result, $\sigma_{xx}^x(\omega, \boldsymbol{q})$ must be zero when $\boldsymbol{q} = 0$ since it flips sign under $\mathcal{M}_x$. In contrast, if $q_x \neq 0$, then $\sigma_{xx}^x(\omega, \boldsymbol{q})$ can be nonzero, which is vividly illustrated in Figure 4b. In Figure 4c, we show how $\sigma_{xx}^x(\omega = 0.1 \text{ eV}; \boldsymbol{q})$ varies with $q_x$. Again, one can see that PLs on the surface and in the bulk have distinct behavior. $\sigma_{xx}^x(\omega, \boldsymbol{q})$ of bulk PLs shows a cubic and monotonic relationship with $q_x$. In contrast, $\sigma_{xx}^x(\omega, \boldsymbol{q})$ on the surface is non-monotonic with $\boldsymbol{q}$, and reverses direction at around $|q_x| \approx 0.6 \text{ nm}^{-1}$. This is somewhat surprising since $|q_x|$ determines the extent to which $\mathcal{M}_x$ is broken, and one may expect that $\sigma_{xx}^x(\boldsymbol{q}, \omega)$ should increase monotonically with $q_x$. However, when $|q_x|$ is large the wavefunction overlap between $(\boldsymbol{k}, E)$ and $(\boldsymbol{k} + \boldsymbol{q}, E + \omega)$ may become smaller, leading to a complicated relationship between $\sigma_{xx}^x(\omega, \boldsymbol{q})$ and $q_x$. As a comparison, we keep $q_x = 0$ and vary $q_y$, and $\sigma_{xx}^x(\omega, \boldsymbol{q})$ remains zero as expected (SM Figure S8), since $q_y$ cannot break $\mathcal{M}_x$.

Finally, we would like to remark that besides inhomogeneous light field, a finite $\boldsymbol{q}$ can also arise from inhomogeneous materials, in the presence of e.g., strain gradient[49] or heterostructures. Inhomogeneous light field and materials should conceptually lead to similar results, although methodologically, the effect of inhomogeneous materials shall be incorporated in a different way.

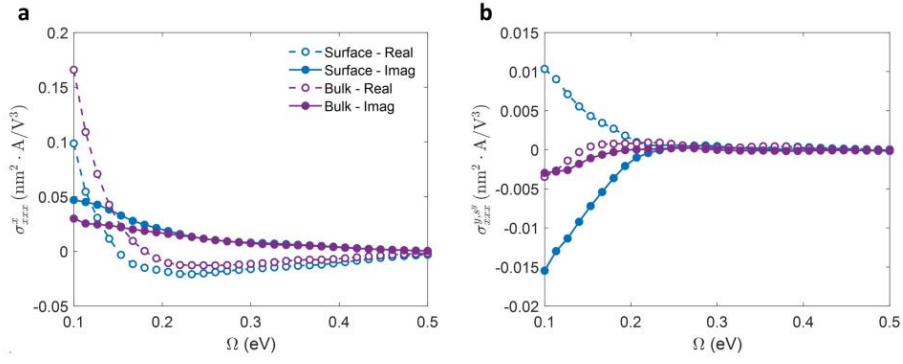

**Figure 5** Third order conductivity of $T_d$-WTe$_2$ for (a) charge current and (b) spin current

**Higher-order response.** Until now, we have been discussing the second-order optical responses, which scales as $\mathcal{E}^2$. In recent years the higher-order responses have also attracted great interest[32,50–53], and efficient high-order responses have been demonstrated separately on the surface[32] and in the bulk[53] of topological materials. Actually, in the same material, the higher-order responses can be distinct on the surface and in the bulk interior as well. Here we consider the third-order response, which can be characterized by the conductivity $\sigma_{bcd}^a(\omega_b + \omega_c + \omega_d; \omega_b, \omega_c, \omega_d)$ – that is, three electric fields $\mathcal{E}_b(\omega_b), \mathcal{E}_c(\omega_c), \mathcal{E}_d(\omega_d)$ with polarizations $(b, c, d)$ and frequencies $(\omega_b, \omega_c, \omega_d)$ are coupled, and a current along direction $a$ with frequency $\omega_b + \omega_c + \omega_d$ is generated. The detailed formula to calculate $\sigma_{bcd}^a(\omega_b, \omega_c, \omega_d)$ can be found in SM Sec. 2.3.

A simple but typical example of the third-order effect is when one applies two laser beams with $\mathcal{E}_b = 2\tilde{\mathcal{E}}_b \cos(\omega t + \phi_b)$ and $\mathcal{E}_c = 2\tilde{\mathcal{E}}_c \cos(2\omega t + \phi_c)$, a static current $j^a$ can be generated as

$$j^a = \tilde{\mathcal{E}}_b^2 \tilde{\mathcal{E}}_c \left[ \sigma_{bbc}^a(0; \omega, \omega, -2\omega) e^{i(2\phi_b - \phi_c)} + \sigma_{bbd}^a(0; -\omega, -\omega, 2\omega) e^{-i(2\phi_b - \phi_c)} \right] \quad (5)$$
$$= \tilde{\mathcal{E}}_b^2 \tilde{\mathcal{E}}_c \operatorname{Re}\left[ \sigma_{bbc}^a(0; \omega, \omega, -2\omega) e^{i(2\phi_b - \phi_c)} \right]$$

where we used $\sigma_{bbc}^a(0; \omega, \omega, -2\omega) = [\sigma_{bbd}^a(0; -\omega, -\omega, 2\omega)]^*$, which is the consequence of time-reversal symmetry. We computed $\sigma_{xxx}^x(0; \omega, \omega, -2\omega)$ with the Green's function formalism, and the results are shown in Figure 5a, where one can see that the third-order responses are distinct for the surface and the bulk as well. Moreover, $\sigma_{bbc}^a(0; \omega, \omega, -2\omega) \equiv e^{i\psi}|\sigma_{bbc}^a(0; \omega, \omega, -2\omega)|$ is a complex number, yielding $j^a = \tilde{\mathcal{E}}_b^2 \tilde{\mathcal{E}}_c |\sigma_{bbc}^a(0; \omega, \omega, -2\omega)| \cos(2\phi_b - \phi_c + \psi)$. Interestingly, since spin flips its direction under time-reversal symmetry, one has the spin current conductivity (Figure 5b) as $\sigma_{bbc}^{a,s^i}(0; \omega, \omega, -2\omega) = -\left[\sigma_{bbc}^{a,s^i}(0; \omega, \omega, -2\omega)\right]^*$, and the spin current obeys $j^{a,s^i} \propto$ $\sin(2\phi_b - \phi_c + \psi_{si})$. Therefore, the magnitude and direction of both charge and spin current can be controlled by the phase difference $2\phi_b - \phi_c$. Furthermore, pure spin current without accompanying charge current can be generated when $2\phi_b - \phi_c + \psi = n\pi + \frac{\pi}{2}, n \in \mathbb{Z}$. These effects are the so-called two-color quantum interference control[54–57].

## Discussions

In this work we used $T_d$-WTe$_2$ as an example to study the surface effects in NLO processes. The reason we choose $T_d$-WTe$_2$ is that the bulk of $T_d$-WTe$_2$ is non-centrosymmetric and semi-metallic, which is similar to the surface from a symmetry or bandgap point of view. Our results indicate that even in this case, the surface responses can still be drastically different from those in the bulk. This is mainly a topological effect. As a comparison, we calculate the PL resolved BPV conductivity of bulk 2H-MoS$_2$ with AA stacking, which is topologically trivial. We find that the surface PL and bulk PL in 2H-MoS$_2$ have almost the same responses under light (SM Figure S7).

As discussed before, in materials with inversion symmetry or nonzero bandgap, the difference between surface and bulk can be more dramatic. In centrosymmetric materials, B(S)PV, as a second-order effect, is forbidden in the bulk interior. Therefore, under light illumination, the currents are purely on the surfaces, where the inversion symmetry is broken. As for topological insulators, the bulk has a finite bandgap and is thus silent under light with below-bandgap frequencies. However, the surface states are gapless, and can be active under light with (in principle) arbitrarily low frequencies. These features are illustrated using Bi$_2$Se$_3$ as an example (SM Figure S6). Note that in Be$_2$Se$_3$, the BPV conductivities on the top and bottom surfaces are opposite, thus the total current shall be zero if the top and bulk surfaces are under the same light field. However, light with above-bulk-bandgap frequencies may not reach the bottom surface if the Be$_2$Se$_3$ is thick enough. In this case, the total current comes solely from the top surface and can be directly used to probe the surface states.

In conclusion, we developed a generic Green's function framework for calculating the NLO properties, which can incorporate many-body effects beyond the single-particle approximation. In future works, we will use this framework to study NLO effects in e.g., strong correlation systems. In this work, the Green's function framework is used to study the surfaces of topological materials, and it is found that under light illumination, the surface can behave disparately from the bulk. Therefore, light can be used to probe the surface effects. On the other hand, when investigating the NLO properties of topological materials, the surface effects shall be carefully considered, particularly when the light penetration depth is small or the material has nanoscale dimensions. For example, in topological (semi-)metals, only tens of PLs (10 ~ 100 nm in thickness) may be active under light, and the surface effect can make a significant difference in the total responses. This is important when using light as a probe of topological properties: the desired bulk properties may be obscured by surface effects, and the experimental results may deviate from theoretical predictions if the surface effects are not properly considered. Also, when searching for materials with large NLO responses for e.g., photodetection or energy harvesting purposes, it may be insufficient to look at only the bulk properties. The true responses can be obtained only when the surface and the bulk are both considered. The surface can act either positively or negatively to the total response and shall be studied case by case.

## Methods

**Green's function formalism.** Electrons in solid-state systems have interactions with e.g., phonons, defects, and other electrons. We call the electron system with these *internal* interactions the *base* system and assume that we have full knowledge of these base systems – we have their Green's functions $G_0$ in hand, which can either be rigorously calculated or be approximately obtained with e.g., perturbative expansions. *External* fields such as light illumination are treated as additional interactions, which can drive the system out of equilibrium. Such external interactions are described by the self-energy terms $\Sigma$. The Green's functions with both internal and external interactions are denoted with $G$, which can be obtained from $G_0$ and $\Sigma$ using Dyson's equation. Regarding the light-matter interaction, one has[58,59] (SM Sec. 1)

$$G^< = \left[\sum_{n=0}^{\infty}(G_0^r \Sigma^r)^n\right] G_0^< \left[\sum_{m=0}^{\infty}(\Sigma^a G_0^a)^m\right] \tag{6}$$

Here $G_0^r/G_0^a/G_0^<$ are the retarded/advanced/lesser Green's function of the base system. $\Sigma^r(\boldsymbol{q}, \omega) = \Sigma^a(\boldsymbol{q}, \omega) = iev \cdot \frac{\mathcal{E}}{\omega} e^{i(\boldsymbol{q}\cdot\boldsymbol{r}+\omega t)}$ is the retarded/advanced self-energy, where $v$ is the velocity matrix, while $\mathcal{E}$ is the electric field with wavevector $\boldsymbol{q}$ and frequency $\omega$. Eqs. (1, 6) provide a systematic and convenient approach for computing the optical responses to an arbitrary order: for the $N$-th order optical response, one simply picks up all terms with $m + n = N$ in Eq. (6). The $\boldsymbol{k}$ and $E$ arguments are omitted in Eq. (6) for simplicity. Note that after each self-energy term $\Sigma^{r/a}(\boldsymbol{q}, \omega)$, the arguments $k$ and $E$ should be shifted by $\boldsymbol{q}$ and $\omega$, respectively, corresponding to the momentum and energy conservation. When the base system is non-interacting, Eqs. (1) and (6) can be reduced to the common single-particle formula for calculating linear and nonlinear optical responses. This is verified theoretically and computationally in the SM Sec. 1.3 and 1.4.

**Ab initio calculations**. The ab initio calculations are based on density functional theory (DFT)[60,61] as implemented in Vienna *ab initio* simulation package (VASP)[62,63]. Generalized gradient approximation (GGA) in the form of Perdew-Burke-Ernzerhof (PBE)[64] is used to treat the exchange-correlation interactions. Core and valence electrons are treated by projector augmented wave (PAW) method[65] and plane-wave basis functions, respectively. For DFT calculations of $T_d$-WTe$_2$, the first Brillouin zone is sampled by a $15 \times 9 \times 3$ $\boldsymbol{k}$-mesh. Then a tight-binding (TB) Hamiltonian is built from DFT results using the Wannier90 package[66] and is used to calculate the NLO responses within the Green's function framework. For the localized NLO responses on each PL, the BZ integration in Eq. (1) is carried out by $\boldsymbol{k}$-mesh sampling with $\int \frac{d\boldsymbol{k}}{(2\pi)^2} = \frac{1}{S}\sum_{\boldsymbol{k}} w_{\boldsymbol{k}}$, where $S$ is the area of the 2D unit cell on each PL and $w_{\boldsymbol{k}}$ is weight factor. For the NLO conductivity of common bulk materials, the $\boldsymbol{k}$-mesh sampling shall be $\int \frac{d\boldsymbol{k}}{(2\pi)^3} = \frac{1}{V}\sum_{\boldsymbol{k}} w_{\boldsymbol{k}}$, where $V$ is the volume of the 3D unit cell. Thus, the conductivities shown in this work differ from the common definition of conductivities by a factor $L$, which is the thickness of the unit cell. For the $\boldsymbol{k}$-integration in Eq. (1), a $\boldsymbol{k}$-mesh of $128 \times 64$ is used for $T_d$-WTe$_2$, while for the $E$-integration, a trapezoidal method is used with an energy interval of 10 meV. The convergence with both $\boldsymbol{k}$ and $E$ integrations are tested.

**Acknowledgment**

This work was supported by an Office of Naval Research MURI through grant #N00014-17-1-2661. The authors acknowledge helpful discussions with Jian Zhou.